# Cavitation-Induced Fusion: Proof of Concept

Max I. Fomitchev-Zamilov

Quantum Potential Corporation, 200 Innovation Blvd, Suite 254, State College, PA 16803

e-mail: max@quantum-potential.com

Cavitation-induced fusion (also known as bubble fusion or sonofusion) has been a topic of much debate and controversy and is generally (albeit incorrectly) perceived as unworkable. In this paper we present the theoretical foundations of cavitation-induced fusion and summarize the experimental results of the research conducted in the past 20 years. Based on the systematic study of all available data we conclude that the cavitation-induced fusion is feasible, doable, and can be used for commercial power generation. We present the results of our own research and disclose a commercial reactor prototype.

## 1. Introduction

Nuclear fusion (which powers the sun) is the energy of the future: 10 microgram of deuterium is equivalent to a barrel of oil. Deuterium is cheap, plentiful and easily extracted from water. Unlike uranium fission in modern nuclear power plants deuterium fusion does not produce radioactive waste. Yet despite 40 years of research and over $20B in government spending (Chu, 2008) on inertial/magnetic confinement projects (ICF/MCF) the fusion power remains out of reach: to this date there are no fusion reactors capable of sustained operation and net energy production. Massive capital expenditures (billions) are necessary to build and maintain ICF/MCF facilities and equally massive technological challenges remain. Because of these difficulties it is prudent to look for other, less costly fusion alternatives.

Cavitation-induced fusion (CIF) is one such alternative. The CIF idea gained popularity when observation of light pulses emitted by collapsing cavitation bubbles revealed unexpectedly extreme conditions within the collapsing bubble cores: temperatures in excess of 30,000K (5 times hotter than the surface of the sun) have been measured directly and even higher temperatures (in the millions degrees K) have been inferred (Flannigan & Suslick, 2010).

As a result an experimental and theoretical work has followed and numerous ideas have been put forward, patents filed (Janssen, et al., 2009) and taken (Putterman S. , 1997), and at least one privately funded company (Impulse Devices) was founded to pursue CIF commercially.

Unexpectedly a misfortune struck this promising field of research. Taleyarkhan and co-authors (Taleyarkhan R. , West, Cho, Lahey, Nigmatulin, & Block, 2002) published what was believed (albeit incorrectly) to be the first successful "bubble fusion" experiment. Their report (which first appeared in Science) with follow-up papers published in Physical Review (Taleyarkhan, Cho, West, Lahey, Nigmatulin, & Block, 2004), stirred a hornet's nest provoking all sorts of nasty developments ranging from academic rivalry, to conflict of interests in research funds appropriation (ICF researchers felt threatened), to tenure and promotion issues and academic misconduct (Krivit, 2011). As a result of the ensuing "bubblegate" scandal Taleyarkhan's career was destroyed (Reich, 2009) and CIF research became a taboo.

What was forgotten amid the outburst of emotions is that cavitation-induced fusion is a *fruitful area of research that must be continued*: no less than 7 independent peer-reviewed reports exist demonstrating neutron emissions from collapsing cavitation bubbles; even heavily criticized experiments by Taleyarkhan's group have been successfully repeated (Xu & Butt, 2005), (Forringer, Robbins, & Martin, 2006), (Bugg, 2006).

Because of potential importance of CIF we have conducted our own feasibility study (Section 4) and completed preliminary work (Section 5) that yielded encouraging results. Therefore, **we propose to conduct a new, thorough experiment that will demonstrate beyond any doubt feasibility of cavitation-induced fusion.** Such experiment is a first step towards commercial net-power producing generator development that may revolutionize the way we generate power opening a path towards green, clean, affordable energy with zero carbon footprint.

## 2. Theoretical Foundations

Under the influence of acoustic waves permeating the liquid tiny dissolved gas bubbles undergo cycles of periodic expansion followed by violent collapse, Fig. 1. The resulting phenomenon is known as cavitation.

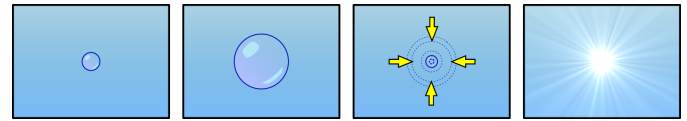

**Fig. 1**. Bubble growth and collapse during cavitation resulting in sonoluminescence (from Wikipedia).

The cavitation bubble collapse can be surprisingly strong – swarms of bubbles can easily eat through metals and destroy machinery – impellers, propellers, rotors, and pipes, Fig. 2.

The reason for this seemingly surprising behavior is that cavitation bubbles act as spherical energy concentrators: total kinetic energy (*E*) of the implosion grows as a cube of the maximum bubble radius $R_{max}$:

$$E = 4/3 \pi R_{max}^3 P_{max} \quad (1)$$

where $R_{max}$ – is the maximum bubble radius and $P_{max}$ is the liquid pressure during the collapse phase (constant pressure is assumed).

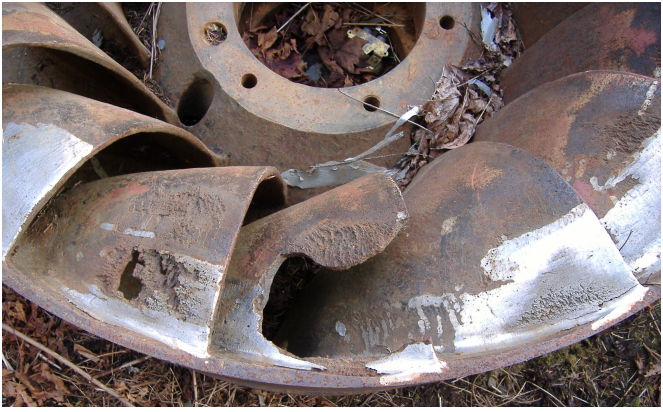

**Fig. 2.** Cavitation damage to a Francis turbine.

What makes this energy concentrating process useful is that this energy can be focused onto a minuscule amount of gas trapped in the initially small (micron-size) gas bubble. From the equation of state for an ideal gas:

$$P_0 V_0 = N k_B T_0 \quad (2)$$

where $P_0$ – initial bubble gas pressure, $V_0 = 4/3 \pi R_0^3$ is the initial bubble volume, *N* – number of atoms of gas in the bubble, $k_B$ – Bolzmann constant, $T_0$ – initial bubble gas temperature, we can estimate maximum energy concentration per atom of gas ($E_a$) as

$$E_a = (k_B T_0)^{-1} (R_{max}/R_0)^3 P_{max}/P_0 \quad (3)$$

Converting the energy into "fusion" units of keV and assuming ambient temperature $T_0 = 300K$ we can rewrite the equation (3) as

$$E_a \approx 4 \times 10^{-5} (R_{max}/R_0)^3 P_{max}/P_0 \; keV \quad (4)$$

Fusion reactions involving deuterium (D/D fusion) occur in meaningful quantities at energies above *100 keV*, while deuterium-tritium (D/T) fusion reactions are going strong in the *10-1,000 keV* range (Fig. 3).

The equation (4) tells us that all we need to do in order to achieve D/T fusion ($E_a = 10 \; keV$) is to attain bubble expansion ratio of $R_{max}/R_0 = 30$ combined with liquid pressure during bubble collapse 10 times in excess of ambient ($P_{max}/P_0 = 10$). While this calculation is very naïve it gives a scope of possibilities.

## 2.1 Rayleigh-Plesset-Keller Equation

More accurate estimate of fusion efficiency can be obtained analytically by solving Rayleigh-Plesset-Keller (RPK) differential equation (Keller & Kolodner, 1956) for cavitation bubble collapse:

$$\left(1 - \frac{R'}{c}\right) \rho R R'' + \frac{3}{2}\left(1 - \frac{R'}{3c}\right) \rho R'^2 =$$
$$\left(1 + \frac{R'}{c}\right)( \; - P_0 - P_d) + \frac{R}{c}(P' - P_d') - \frac{2\sigma}{R} - \frac{4\eta R'}{R} \quad (5)$$

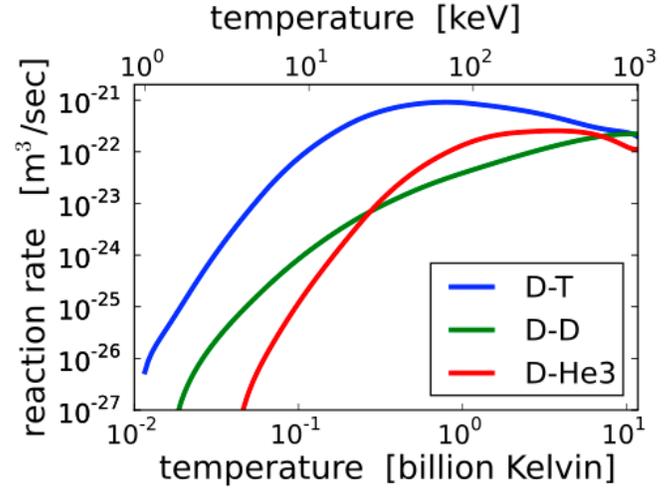

**Fig. 3.** Dependence of deuterium-tritium (D/T), deuterium-deuterium (D/D) and deuterium-helium3 fusion reaction rates on temperature and energy (*1 keV ≈ 10,000,000K*).

where *R* is bubble radius, *R'* and *R''* are first and second derivatives of *R*, *P* is gas pressure within the bubble and *P'* is its derivative, $P_d$ is acoustic driving pressure and $P_d'$ is its derivative, *c* is the velocity of sound in the liquid, *ρ* is liquid density, *σ* is liquid surface tension, and *η* is dynamic viscosity of the liquid.

The RPK equation (5) accounts for liquid viscosity ($\frac{4\eta R'}{R}$ term) and surface tension ($\frac{2\sigma}{R}$ term) as well as for losses due to shockwave formation because of liquid compressibility ($\frac{\_'}{c}$ terms).

In order to characterize the conditions within the collapsing bubble and obtain the functions *R = R(t)* and *P = P(t)* the RPK equation must be solved numerically together with the equation of state for the bubble gas. The resulting solution is quite sensitive to the choice of the equation of state during the last stage of collapse (which is the most interesting stage from the standpoint of nuclear fusion).

Ignoring for the time being the last crucial stage of collapse, adiabatic equation of state:

$$PV^\gamma = const \quad (6)$$

where *γ* is the ratio of specific heats (*γ = 5/3* for monoatomic and *γ = 7/5* for diatomic gas), provides nice fit for experimental data, Fig. 4.

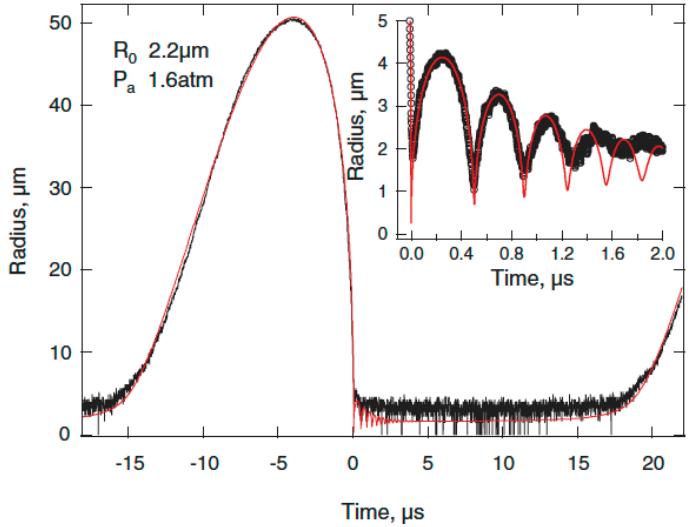

**Fig. 4.** Analytical solution to Rayleigh-Plesset equation compared to experimental data, from (Bass, Ruuth, Camara, Merriman, & Putterman, 2008). $P_a$ refers to maximum driving acoustic pressure that changes as $P_d(t) = P_a \sin(2\pi f t)$.

## 2.2 Deuterium Equation of State

An accurate equation of state for deuterium can be given as a system of functions of pressure $P$ and energy $\varepsilon$ (Moss, Clarke, White, & Young, 1996):

$$P = R'T\rho[1 + m_D(1 + 2m_I)] + \frac{E_c \rho_0}{1 - \frac{3}{n}}\left[\left(\frac{\rho}{\rho_0}\right)^{\frac{n}{3}+1} - \left(\frac{\rho}{\rho_0}\right)^2\right]$$

$$\varepsilon = \left(\frac{5}{2}R'T + \frac{R'\Theta}{e^{\Theta/T}-1}\right)(1 - m_D) + m_D R'T_D + \frac{3}{2}R'T(2m_D)(1 + m_I) + 2m_D R'^{m_I T_I} + \frac{E_c}{\frac{1}{3}n-1}\left[\left(\frac{\rho}{\rho_0}\right)^{\frac{n}{3}} - \frac{1}{3}\left(\frac{\rho}{\rho_0}\right)\right] + E_c \quad (7)$$

where $R' = R/M(D_2)$ is the gas constant for deuterium ($R = 8.3$ J/(K mol), $M(D_2) = 4.03$ g/mol – molecular deuterium molar mass), $T$ – temperature, $\rho$ – is deuterium density ($\rho = m_0/V$, $V = 4/3\,\pi R^3$, $m_0$ – mass of bubble gas, which is assumed to be constant), $m_D$ – fraction of disassociated deuterium ($0 \leq m_D \leq 1$; $T_D = 4.5$ eV), $m_I$ – fraction of ionized deuterium ($0 \leq m_I \leq 1$; $T_I = 13.6$ eV), $E_c = 1.09 \times 10^6$ J/kg – binding energy, $\rho_0 = 202$ kg/m$^3$ – solid deuterium density at $0K$, $n = 5$, $\Theta = 4394K$ – accounts for vibrational energy contribution.

Both disassociated and ionized deuterium fractions can be approximated as (Moss, Clarke, White, & Young, 1996):

$$m_k = 0.5\left[\tanh\left(\frac{7(T - 0.9T_k)}{T_k}\right) + \tanh(6.3)\right] \quad (8)$$

where $k$ is either $D$ or $I$.

Although the equation of state (7) accounts for rotational energy, molecular disassociation, ionization, and inter-molecular forces, we nevertheless assume that:

a) There is no mass exchange between bubble gas and the liquid;

b) There is no heat exchange between bubble gas and the liquid;
c) There is no shockwave formation within the bubble gas.

While the first assumption can be made true by selecting a liquid with low vapor pressure and/or high vapor accommodation coefficient and the second assumption can be satisfied when the collapse is very rapid (the bubble gas has no time to achieve thermal equilibrium with the bubble wall), the third assumption breaks down when the bubble collapse reaches supersonic velocities. In the same time shockwaves are known to form within the collapsing bubbles (Wu & Roberts, 1993) and thought to play important role in sonoluminescence. Hence any numerical solution to RPK equation assuming adiabatic compression (as well as uniform bubble pressure and temperature) will yield underestimated peak temperature and overestimated average pressure and therefore should be considered a *lower bound* for actual peak temperature and peak pressure that will occur within the collapsing bubble due to shock formation.

## 2.3 Neutron Production Estimation

The fusion rate (fusions per second) $f$ can be calculated as:

$$f = \frac{1}{2}\left(\frac{\rho}{M}\right)^2 \sigma v(T) V \quad (9)$$

where $\sigma v(T)$ – is reaction cross-section.

For D/D and D/T fusion the reaction cross-section can be approximated as:

$$\sigma v(T) = C_1 \zeta^{-5/6} \xi^2 e^{-3\zeta^{1/3}\xi} \quad (10)$$

where

$$\zeta = 1 - \frac{C_2 T + C_4 T^2 + C_6 T^3}{1 + C_3 T + C_5 T^2 + C_7 T^3} \quad (11)$$

$$\xi = \frac{C_0}{T^{1/3}} \quad (12)$$

and $C_0 - C_7$ are constants.

Then the total fusion yield per bubble collapse $N$ is

$$N = \int_0^\tau f(t)\,dt \quad (13)$$

Numerically solving the equations (5) and (7) together we obtain the functions of bubble radius $R$ (from which we can calculate $\rho$) and bubble temperature $T$ that we need to estimate the fusion yield $N$ using the equations (9) – (13).

For example, a low-pressure ($P_0 = 0.01$ Pa) bubble in mercury composed of atomic D/T mixture compressed from maximum radius $R_{max} = 100$ μm by driving pressure $P_a = 100$ bar will reach peak temperature $T_{max} \approx 80{,}000{,}000K$ and peak pressure $P_{max} \approx 2 \times 10^{12}$ Pa thus producing $N \approx 36$ fusion reactions per collapse (max bubble wall velocity is ~12 km/s or Mach 8).

Another example, a 10-micron D/T bubble in liquid tungsten blown to a rather large maximum radius $R_{max} = 7$ mm and then compressed by $P_a = 1000$ bar driving pressure will result in peak temperature $T_{max} \approx 110,000,000K$ and an astonishing $N \approx 2.7 \times 10^{11}$ fusion reactions (yet only ~0.9% of the bubble kinetic energy is spent on heating gas with the remaining *99.1%* of it lost due to liquid compressibility on acoustic wave radiation). In the latter case the final temperature spike lasts less than *0.1 μs*).

Once again these estimates represent lower bounds since the calculation was done assuming adiabatic compression and uniform temperature and pressure within the bubble, which will not be the case during final supersonic stage of collapse.

More realistic calculation involves hydrodynamic (Moss, Clarke, White, & Young, 1996), (Nigmatulin, et al., 2005) or molecular dynamic (Bass, Ruuth, Camara, Merriman, & Putterman, 2008) simulation. Using the most accurate equation of state for molecular deuterium available, Moss simulated 90-micron deuterium bubble collapse under in water. To produce shock the driving sinusoidal pressure was spiked with *5-bar* pressure pulse. As a result they predict 2.5 D/D fusion events per hour assuming metronomic bubble collapse at *27.6 kHz* frequency, Fig. 5 (the estimate for D/T mixture would have been ~100 higher due to proportionally larger D/T reaction cross-section).

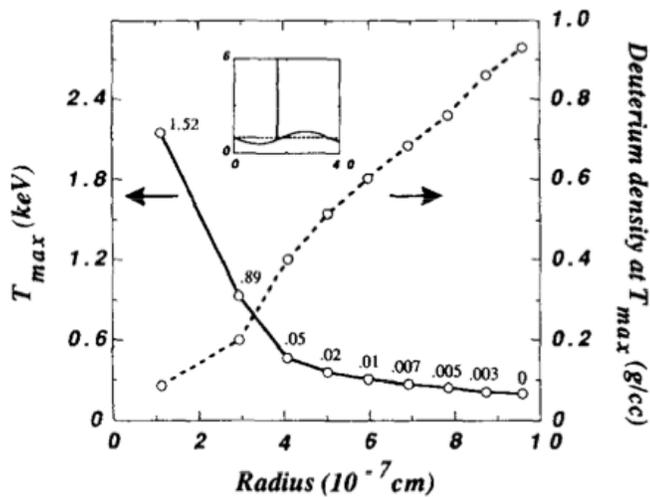

**Fig. 5.** Calculated peak temperature (solid line) and density (dashed line) at the centers (circles) of the innermost hydrodynamic modeling zones within the collapsing bubble cores.

An even more detailed analysis was conducted by (Nigmatulin, et al., 2005) for the case of acetone where an unprecedented effort was spent on deriving the equation of state for this liquid and accounting for all possible effects occurring during bubble collapse (including disassociation, ionization and shock formation). Their simulation and analysis clearly reveals shockwave formation and peak temperatures well in excess of 100 million K, which result in neutron yield of 12 neutrons/collapse for D/D fusion. Other important conclusions of Nigmatulin's team are:

a) The final stage of collapse is so fast that acetone does not have time to disassociate (i.e. no energy is wasted on endothermic chemical reactions);
b) The collapse is so fast that electrons do not have time to thermalize and thus do not contribute to pressure (this results in more violent shockwave due to lower plasma pressure);
c) Bubbles in cluster experience much stronger shocks due to individual bubble shockwave interactions (Nigmnatulin, 1991).

Similar intriguing results were obtained via molecular dynamics simulation of rapidly collapsing bubbles in water ($R_o = 2$ *μm*, $R_{max} = 55$ *μm*, $P_a = 1.6$ *bar*), which revealed mass segregation and strong shockwave formation in gas mixtures comprised of xenon and helium (Fig. 6) with peak temperatures in the range of 10 to 100 million K – Fig. 7 (Bass, Ruuth, Camara, Merriman, & Putterman, 2008). The lower bound of neutron production was estimated at $6 \times 10^{-5}$ neutrons/collapse.

We too have conducted our own simulations using molecular dynamics software initially developed by Bass. For the case of 5% D/T mixture in 95% mercury vapor for the hard-sphere model the fusion rate was ~20,000 reactions per collapse with peak temperatures over 100 million K.

Thus, the modeling results for realistic bubble conditions are very encouraging. However, experimental verification of peak temperatures and pressures by capturing and analyzing spectra of sonoluminescence light flashes proved difficult due to opaque plasma formation in bubble core (Flannigan & Suslick, 2010), which means that no light can escape superheated bubble core during the final stage of collapse. Nevertheless, temperatures in excess of 20,000 K and probably as high as 1 million K have already been measured (Flannigan & Suslick, 2010).

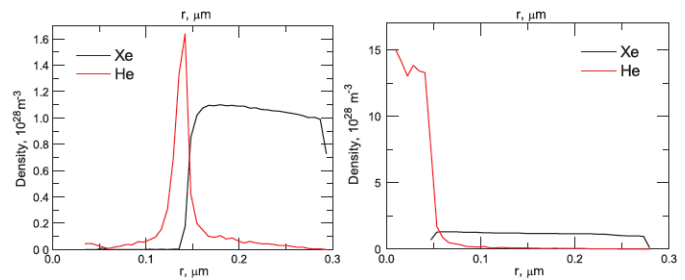

**Fig. 6.** Two consecutive stages of bubble collapse illustrating the compacting/segregating effect of heavy gas shockwave (xenon, black line) on light gas (helium, red line), from (Bass, Ruuth, Camara, Merriman, & Putterman, 2008).

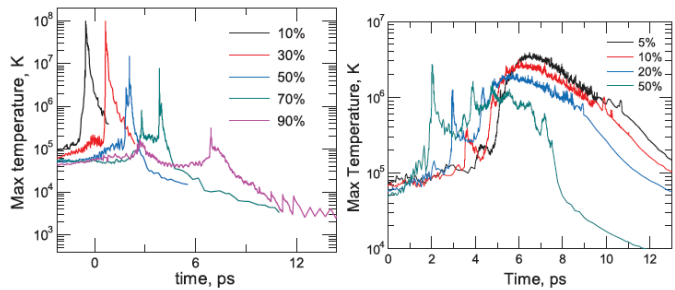

**Fig. 7.** Helium temperatures in the bubble core for hard sphere (left) and variable soft sphere (right) molecular dynamics simulation of bubble implosion, $R_0/R_{max} = 30$, from (Bass, Ruuth, Camara, Merriman, & Putterman, 2008).

## 3. Experimental Results

### 3.1 Lipson et al. (USSR)

Contrary to popular belief the earliest cavitation-induced fusion work commenced not at Oak Ridge National Laboratory (USA) or Purdue University but in the USSR in the early 1990s. E.g. in 1990 Lipson and co-authors used titanium vibrator to cavitate heavy water, there experimental setup is shown on Fig. 8. The mechanism of fusion in their case involved titanium deuteride (TiD) layer formation on the surface of the vibrator (Lipson, et al., 1990), which was violently pierced and compacted by heavy-water microjets formed due to near-surface collapse of cavitation bubbles at the vibrator tip, Fig. 8 (inset). They have also observed neutron emission when intermetallic $LaNi_5D_6$ powder was dispersed in heavy water due to bubble microjet focusing on deuterium-reach microparticles with high surface area. The measured neutron flux was ~ 1 n/s or ~30 times in excess over background of 0.035 n/s.

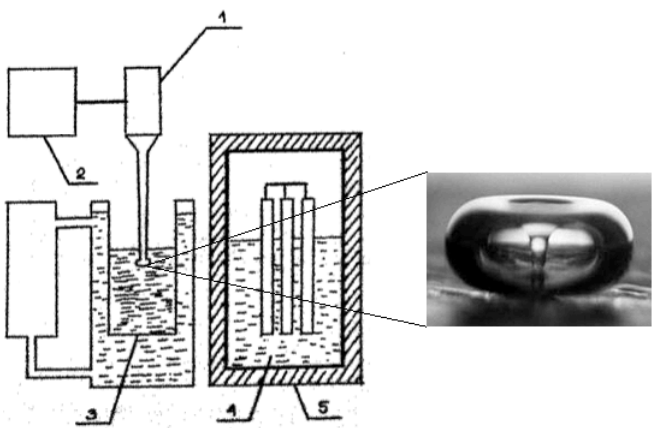

**Fig. 8.** Experimental apparatus used by (Lipson, et al., 1990): 1 – titanium vibrator; 2 – ultrasonic generator; 3 – vessel with $D_2O$, 4 – oil-filled neutron detector with three proportional counters, 5 – cadmium shielding. Inset: cavitation bubble near vibrator surface featuring a microjet impacting on the surface. These microjets are known for their strength and are a chief reason behind cavitation erosion.

### 3.2 Bityurin et al. (Russia)

Subsequent work was carried out by Bityurin and co-authors at the Joint Institute for High Temperatures of the Russian Academy of Sciences (Bityurin, Bykov, Velikodny, Dyrenkov, & Tolkunov, 2008). The group studied the effect of shockwaves on deuterated liquid ($D_2O$) with high (20-95%) bubble content. Their experimental setup is shown on Fig. 9 and includes admission of deuterium bubbles into deuterated liquid and crushing them with a shockwave generated via explosion of a semicircular wire (20) due to high current pulse. The resulting shockwave propagates in the bubble/liquid phase and focuses much stronger than in the pure liquid due to shockwave amplification effects in the gaseous phase. The observed shockwave amplification is somewhat similar to the pressure enhancement observed in bubble clusters (Brennen, 1995). The group used Indium (beta-decay) detectors to measure neutron flux and estimate total neutron yield at $10^8$-$10^{10}$ per explosion.

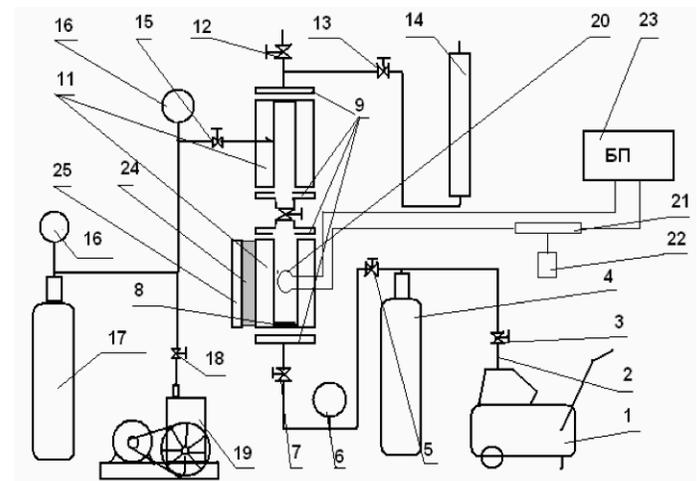

**Fig. 9.** Experimental apparatus used by (Bityurin, Bykov, Velikodny, Dyrenkov, & Tolkunov, 2008). 1 – compressor, 2 – pipes, 3 – valve, 4 – receiver, 5 – valve, 6 – pressure gauge, 7 – valve, 8 – porous titanium disperser, 9 – aluminum casing, 10 – valve, 11 – hollow Plexiglas cylinders, 12 – control valve, 13 – valve, 14 – gas flow meter, 15 – valve, 16 – vacuum gauge, 17 – receiver, 18 – valve, 19 – vacuum pump, 20 – exploding wire, 21 – discharge switch, 22 – switch control, 23 – power source, 24 – neutron mediator, 25 – Indium sheet (neutron detector).

### 3.3 Smorodov et al. (Russia)

Yet another particularly elegant experiment was conducted by (Smorodov, Galiakhmetov, & Il'gamov, 2008). Smorodov and co-authors were able to achieve extreme energy concentration by creating very large ($R_{max} \approx 3\ mm$ in diameter) deuterium bubbles in glycerin and crushing them with high impact force creating equivalent pressures in the excess of $P_a \approx 1,000\ bar$. Plugging these numbers in equations (4) and (5) yields astonishing energy concentration factors, supersonic implosion and strong shockwave formation in the bubble gas. Smorodov et al. used calibrated helium-3 neutron detectors with adjustable signal rejection and their results are shown on Fig. 10.

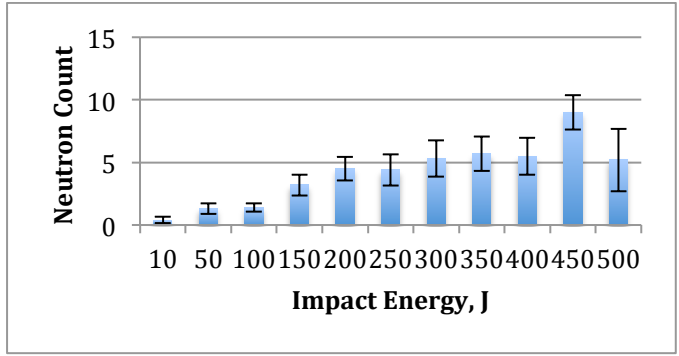

**Fig. 10.** Average (over 10 trials) neutron counts per bubble collapse as a function of impact energy from (Smorodov, Galiakhmetov, & Il'gamov, 2008). Impact energy of 450 J corresponds to peak pressure of ~1,000 bar. Each experiment (individual bar on figure) took 110 seconds to execute, which corresponds to no more than 9 cumulative background neutrons per trial or ~1 background neutron per bubble collapse on average. Therefore the measured neutron flux from bubble collapse for the impact energy of 450 J is 9 times over background.

### 3.4 Taleyarkhan et al. (ORNL)

The results of the Russian experiments were published in top-tier Russian peer-reviewed journals and were not translated into English until recently and therefore are not widely known. In the same time much better publicized work (which is incorrectly credited with the discovery of cavitation-induced fusion) was performed by Taleyarkhan and his colleagues (Taleyarkhan R., West, Cho, Lahey, Nigmatulin, & Block, 2002), and their experimental setup is shown on Fig. 11.

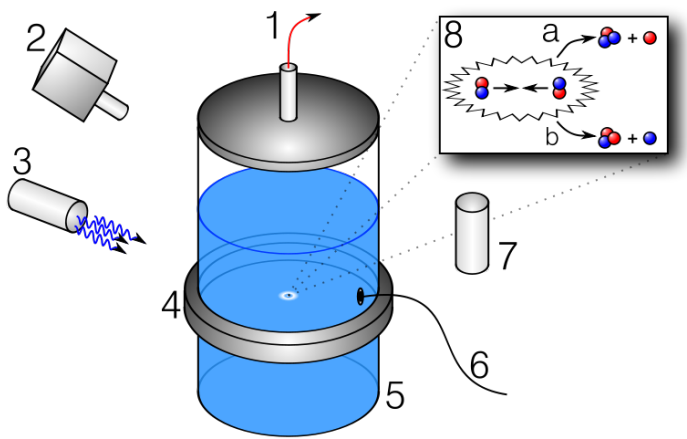

**Fig. 11.** CIF device used by (Taleyarkhan R., West, Cho, Lahey, Nigmatulin, & Block, 2002): 1 - vacuum pump; 2 - liquid scintillator; 3 - neutron source; 4 - acoustic wave generator; 5 - test chamber with fluid; 6 – microphone; 7 - photomultiplier tube; 8 - two deuterium atoms collide; 8a - possible fusion event creating helium and a neutron; 8b - possible fusion event creating tritium and a proton.

Then at Oak Ridge National Laboratory (ORNL), Taleyarkhan et al. constructed a very precise glass resonator (5) vibrated by piezoelectric cylinder (4) mounted on the resonator's outer surface. The resonator cavity was filled with chilled ($0^oC$) and well-degassed deuterated acetone (acetone-D6). Piezo-amplifier was driving the transducer at $f = 19.3\ kHz$ creating peak pressures of $P_a = 15\ bar$ in the resonator's center. Because acetone was well degased an external neutron source was required to seed the cavitation bubbles. Talryarkhan et al. used pulsed neutron generator (PNG) synchronized with pizeo-amplifier such that neutrons were emitted when the pressure within the resonator was at its minimum (i.e. maximum liquid tension).

Talryarkhan's experiment was a huge success as they observed neutron emission well in excess of natural and PNG residual background coinciding with the bubble cluster collapse; the spectrum of the detected neutrons was consistent with the D/D fusion and tritium production was also observed (Taleyarkhan, Cho, West, Lahey, Nigmatulin, & Block, 2004).

Taleyarkhan's unexpected success created a lively discussion as well as fierce criticism (Shapira & Saltmarsh, 2002), (Putterman, Crum, & Suslick, 2002) which, however, was addressed in subsequent publications by Taleyarkhan's group (Taleyarkhan, et al., 2008).

### 3.4 Taleyarkhan et al. (Purdue)

Later (already at Purdue) Taleyarkhan et al. staged a different experiment, in which PNG was eliminated and replaced with alpha-particle source (uranium nitrate sault) dissolved in acetone-D6 enriched solution (Taleyarkhan R., West, Lahey, Nigmatulin, Block, & Xu, 2006). The neutron yield in this experiment was lower because alpha particle emission was random and could not be synchronized with the maximum tension in the liquid. Nevertheless the observed neutron flux was well above natural background and there was not a neutron source involved in the experiment that could have confused the results.

### 3.5 Xu, Forringer, Bugg (USA)

It should be mentioned, that Talerarkhan's experiments were repeated by other parties, specifically by (Xu & Butt, 2005), (Forringer, Robbins, & Martin, 2006), and (Bugg, 2006). However, the replications were not quite as independent as the scientific community would have liked (Xu was a former student of Taleyarkhan's and Forringer and Bugg repeated the experiments at Taleyarkhan's lab while visiting Purdue). The only truly independent replication effort that was published in peer-reviewed literature was conducted at UCLA and was unsuccessful (Camara, Hopkins, Suslick, & Putterman, 2007), in part because the replicating team failed to fill the resonator with acetone all the way to the top reflector and injected incondensable gas in their system (Lahey, 2011, private communication).

Unfortunately, despite solid results and because of the scandal at Purdue that involved Taleyarkhan and former head of School of Nuclear Engineering Lafteri Tsoukalas, Taleyarkhan's results are largely dismissed by scientific community without due consideration while equally impressive results published in Russian peer-reviewed literature are for the most part unknown. Nevertheless, the analysis of all available *peer-reviewed* literature on the subject points to an unmistakably nuclear phenomenon that has been demonstrated repeatedly for the past 20 years. In other words, *cavitation-induced fusion is real*.

## 4. Commercial Reactors

### 4.1 Feasibility

While interesting from purely scientific point of view, cavitation-induced fusion has immediate practical application in commercial power generation and heating. The earliest feasibility study of "bubble fusion" dates back to 1995 and was conducted at Los Alamos National Laboratory by Krakowski (Krakowski, 1995). Krakowski thoroughly mapped the process parameter space and concluded that cavitation-induced fusion is feasible and must be developed further. Process-wise the most important conclusion of (Krakowski, 1995) is that slow (isothermal) expansion and fast (adiabatic) collapse of bubbles is necessary for CIF to function (such slow expansion and fast collapse is typically accomplished by ordinary sinusoidal acoustic drive).

Here we present our own feasibility estimate. A commercial reactor will comprise a large volume of carrier liquid saturated with multiple fuel-rich bubbles undergoing periodic expansion and collapse, which can be initiated acoustically (i.e. via piezoelectric transducers), hydrodynamically (e.g. by passing carrier liquid through a system of orifices), or mechanically (via mechanical action of a piston or a hydraulic press). New bubbles will be periodically injected to replenish depleted fuel and old bubbles will be recycled. As a result of the reactor operation the carrier liquid temperature will increase and the excess heat must be carried away via a heat exchanger. The so-obtained heat can be used to power a steam generator/turbine to generate AC power. Direct conversion of nuclear fusion products into electric power (i.e. bypassing heat exchanger, turbine and generator) is possible when special nuclear fuel (such as boron-hydrogen mixture) is used to produce neutron-free reactions. However, proton-boron cycle and other similar reactions generally require higher temperatures and thus outside the scope of this proposal. Conventional D/T or D/D mixture will produce neutrons and heat that is best harvested via conventional heat exchangers and thus will require steam turbine to drive a dynamo producing A/C power.

Regardless of design we can write total reactor power output *W* as follows:

$$W = V_{liquid}\, \rho_{bubble}\, E_{bubble}\, f \qquad (14)$$

where $V_{liquid}$ – total volume of carrier liquid, $\rho_{bubble}$ – cavitation bubble density, $E_{bubble}$ – energy production per bubble, $f$ – bubble collapse (driving pressure) frequency.

Energy per bubble $E_{bubble}$ can be expressed as:

$$E_{bubble} = N\, E_R \qquad (15)$$

where $N$ – is the number of fusions per bubble collapse and $E_R$ – is energy per reaction.

From the stand point of the reactor efficiency the most desirable nuclear fuel is 50/50 deuterium/tritium mixture as D/T fusion reaction has the highest cross section and thus is most easily achieved. D/T fuel is also the most desirable from the standpoint of power output as D/T fusion results in the most energy per reaction:

$$D + D = T\ (1\ MeV) + p\ (3\ MeV)$$
$$D + D = {}^3He\ (0.8\ MeV) + n\ (2.5\ MeV) \qquad (16)$$
$$D + T = {}^4He\ (3.5\ MeV) + n\ (14\ MeV)$$

The runner up is D/D reaction, which has ~100 times lower cross section (Fig. 3) and produces ~4 times less energy ($E_R \approx 4\ MeV$ for D/D vs. $E_R \approx 17.5\ MeV$ for D/T).

From the equation (14) it follows that to boost energy output per unit of volume we must maximize bubble density and collapse frequency. From RPK equation (5) it follows that under strong acoustic drive conditions ($P_d \geq 100\ bar$) smaller $R_{max} < 100$-*micron* bubbles can oscillate much faster than larger mm-size bubbles: frequencies as high as *150 kHz* are possible for smaller bubbles vs. *20-30 kHz* for larger bubbles. However, due to their relatively small size and modest expansion ratio the micron-size bubbles cannot produce as many fusions as mm-size bubbles.

From the equations (14) and (15) we can express the requirement for the number of reactions per bubble as:

$$N = \Omega/(\rho_{bubble}\, E_R\, f) \qquad (17)$$

where $\Omega$ – is power density:

$$\Omega = W/V_{liquid} \qquad (18)$$

Bubble density can be approximated as:

$$\rho_{bubble} = (X\, R_{max})^3 \qquad (19)$$

where $X$ is average distance between bubbles in the liquid in terms of maximum bubble radius.

Assuming D/T fuel, power density of 10 kilowatt per liter, and *100-micron* bubbles spaced out 10 maximum radii from each other ($X = 10$, $\rho_{bubble} = 10^9$ *bubbles/liter*) and driving frequency of *140 kHz* the resulting requirement of fusions per bubble is $N \approx 25$. To verify that such requirement is realistic we must pick a liquid and driving conditions and numerically solve the equations (5) and (7) and estimate neutron production using the equations (9) – (13). It turns out, that a low-pressure $R_{max}$ = *100-micron* D/T bubble in mercury driven at $P_a$ = *100 bar* will produce $N \approx 36$ fusions per collapse thus satisfying our requirement.

Increasing bubble density by reducing maximum bubble size from $R_{max}$ = *100-micron* to $R_{max}$ = *70-micron* requires only $N \approx 9$ fusions per collapse whereas analytical calculation yields $N \approx 6$ fusions per collapse for the *70-micron* mercury bubble.

Substituting mercury for liquid tungsten boosts the reaction yield to $N \approx 150$ fusions per collapse under the same condition (17 times more than required for our target power density).

## 4.2 Bubble Cluster Effects

Note that our fusion estimates correspond to lower bounds as we assume adiabatic collapse with uniform density and pressure and neglect shockwave formation in the bubble gas. We also ignored bubble cluster effects (Nigmnatulin, 1991), (Brennen, 1995), which results in much higher driving pressure in the cluster center due to high shockwave pressure $P_{shock}$ of rebounding bubbles:

$$P_{shock} \approx 100 \, R_{max} \, P_d/r \qquad (20)$$

where *r* is distance from the rebounding bubble.

Thus we can effectively expect 10-times higher driving pressure than we used in our calculations above to estimate reaction yield per implosion.

## 4.3 Spherically Symmetric Collapse

We must be note, that our estimates are based on tacit assumption of spherically symmetric collapse, which we expect to be violated at high driving pressures. To evaluate this problem (Nagrath, Jansen, Lahey, & Akhatov, 2006) conducted a simulation of bubble collapse using fluid dynamics and final element method. Their modeling predicts ellipsoidal bubble shape during the final supersonic stage of collapse – fig. 12. They also find that despite the ellipsoidal bubble shape pressure distribution within the bubble core remains spherical. Furthermore, high surface tension and high viscosity act as damping forces against bubble shape instabilities. Therefore bubbles in liquid metals (which have highest surface tension of all known liquids, e.g. $\sigma$ = *2300 N/m* for liquid tungsten vs. $\sigma$ = *70 N/m* for water) will remain spherical notwithstanding extreme driving pressures of $P_a$ = *1000 bar*.

## 4.4 Reactor Drive Power Requirements

For a commercial CIF reactor to be net-power producing the reactor must produce more power than is required to operate it**.** This is one of the problem that is still unresolved for conventional MCF/ICF fusion megaprojects: existing prototypes of these reactors consume more energy to power megawatt lasers and huge magnets than they are able to harvest as a result of fusion reactions.

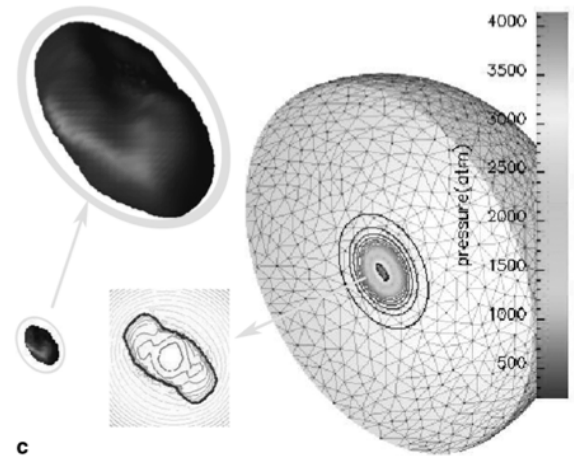

**Fig. 12.** Final supersonic stage of bubble collapse from (Nagrath, Jansen, Lahey, & Akhatov, 2006). Note that despite ellipsoidal overall bubble shape pressure distribution within the bubble core is still spherical.

Clearly, CIF reactor will require power to operate: the power is needed to produce pressure driving the bubble expansion and collapse. (Krakowski, 1995) has considered this problem in great detail and concluded that net-power from CIF is possible. Here we give our own considerations:

1) In the case of a liquid-metal CIF reactor no power needs to be spent on maintaining liquid temperature except for the cold-start because:
   a. Heat is a byproduct of fusion
   b. Heat is a byproduct of inefficient drive
   c. Appropriate thermostatic conditions will be implemented to maintain the liquid pool temperature even if the power is turned off.
2) Most power will be consumed on production of driving pressure:
   a. Piezoelectric transducers are only about 10% efficient;
   b. Hydrodynamic or mechanical drive may be more efficient but unlikely to be more than 20-30% efficient (typical engineering efficiency);
   c. Acoustic drive must be implemented in a high-Q resonant chamber to minimize losses;
   d. In all cases power loss translates into heating of carrier liquid, and this heat is not lost but removed via heat exchanger and utilized for power generation;
3) Electric generators are typically 97% efficient, however we will assume 50% power loss due to steam generation and turbine hardware inefficiency.

To estimate power requirements to drive a bubble cluster of $V_{liquid}$ = *1 liter* of volume at $P_a$ = *100 bar* it is sufficient to apply $P_a$ = *10 bar* of peak pressure and rely on bubble cluster pressure amplification mechanism – equation (20) – to increase the peak pressure tenfold. Then the impact energy necessary to drive the process can be estimated as:

$$E_{impact} = k\, \Delta x^2/2 = mgh \qquad (21)$$

where $k$ is the bulk modulus of the liquid ($k = 28.5 \times 10^9\ Pa$ for mercury) and $\Delta x$ is the compaction displacement of the liquid (compression of a liquid is analogous to that of a spring), $m$ – weight mass, $g$ – free fall acceleration, and $h$ – weight fall height.

The peak impact pressure $P_{peak}$ arises from the max reaction force of the compressed liquid:

$$P_{peak} = F_{max}/S = k\, \Delta x/S = \frac{1}{S}\sqrt{2kE_{impact}} \qquad (22)$$

where $S$ is the area being impacted.

From the equation (22) it follows that to create $P_{peak}$ = *10 bar* impact pressure in 1L of liquid mercury contained in 10-cm tall cylinder we need an impact energy $E_{impact} \approx 0.002\ J$.

In the same time total power production per impact due to D/T fusion is:

$$E_{fusion} = N\, E_R\, V_{liquid}/\rho_{bubble} \qquad (23)$$

Assuming $N$ = *10* fusions per bubble and bubble density of $\rho_{bubble}$ = *$10^9$ bubbles/litre* (consistent with our previous computations) we obtain $E_{fusion}$ = *0.028 J*. Thus, $E_{fusion} \approx 14\, E_{impact}$ and the power generation process is clearly feasible.

## 4.5 Fusion Process Optimization

Exponential dependence of fusion yield on temperature and cubic dependence of total power on maximum bubble radius makes CIF process easy to optimize: e.g. an increase in temperature of just few percent will double the power (a 10-fold bubble core temperature increase results in astonishing 500,000 power boost). **Hence even minor process improvements (such as bubble gas pressure reduction or driving pressure increase) will result in exponential increase in power and efficiency.**

In this regard CIF process is unique because it does not abide by the law of diminishing returns that plague typical engineering problems where massive effort is required in order to achieve just a few percent of efficiency increase. Quite on the contrary, CIF is an engineers dream because minor process improvements results in huge efficiency boosts.

## 4.6 Fusion Reactor Safety

Reputation of nuclear power was seriously tarnished due to recent Fukushima disaster forcing nations (e.g. Germany) to abandon their nuclear plans. We wish to emphasize that *nuclear fusion* unlike conventional *nuclear fission* is clean and green technology that does not produce radioactive waste. Therefore nuclear fusion power generation is as clean and as safe as solar power. Granted, D/D and D/T fusion produces neutron radiation that is harmful to humans. However, this radiation is easily screened by hydrogen-rich shielding such as polyethylene or water. Moreover neutron emission stops as soon as reactor is shut down and it the future it should be possible to design reactors operating on neutron-free (e.g. proton-boron) cycle.

Another important aspect of CIF reactors is inherent safety against runaway "chain reactions". If for whatever reason reactor temperature increases and the excess energy is not transported away from the reactor the reaction yield will plummet due to excessive vapor formation in the cavitation bubbles. All discussion in this proposal assumed very low (nearly-zero) vapor pressure, which is possible to achieve with heavy organic liquids and liquid metals for a certain rather narrow range of temperatures (vapor pressure grows exponentially with temperature). Therefore if a reactor operation mode is skewed towards higher power output the bubble gas temperature will rise producing more vapor, which will rapidly quench fusion by increasing the mass of the gas subjected to compression work. In other words, should a reactor fail (e.g. due to heat exchanger damage in a natural disaster) the reactor will automatically and quickly shutdown without catastrophic explosion. This is yet another advantage of CIF over nuclear fission: while fission reactors require constant maintenance to remain cool (hence Fukushima disaster that resulted in reactor core meltdown when backup cooling generators failed), CIF reactors *will not operate* unless heat is constantly removed from the system and will automatically shutdown when the power production exceeds engineered parameters.

## 4.7 Fusion Fuel Considerations and Tritium Safety

From the standpoint of power output and efficiency 50/50 deuterium/tritium mixture is the most desirable nuclear fuel. Tritium is a radioactive isotope of hydrogen with half-life of 12.3 years. Tritium is only mildly radioactive and beta-decays into helium-3. Beta radiation does not persist and easily screened and mitigated by common materials (a sheet of paper will stop beta radiation, which is nothing more than a flux of high-energy electrons). In fact despite its radioactive nature tritium is routinely used for illumination (tritium vials) and is harmful only when inhaled or ingested directly in *substantial* quantities.

At the moment of writing tritium costs $30k per gram (Willms, 2003). Despite its high costs tritium is universally touted as fusion fuel for conventional ICF/MCF megaprojects with the expectation that the tritium costs will come down when mass production of the substance begins to supply the fusion industry. Right now tritium is produced as a byproduct of nuclear reactions at research facilities and available quantities are therefore minuscule. At the current price power generation via D/T fusion will cost $0.30/kwh, or 4.5 times higher than the current cost of electric power generation and transmission of $0.07/kwh. Still, even at this higher-than-electric cost D/T fusion is commercially attractive for trucking and air transportation where the price of diesel fuel and gasoline is a major cost factor.

Fortunately, CIF process can operate on pure deuterium fuel, which is much cheaper, abundant, and non-radioactive. When compared to D/T, D/D fusion is harder to achieve due to ~100 times lower reaction cross-section and the power output is 4.5 times. Still, D/D fusion has been already demonstrated in a number of CIF experiments and therefore is in principle feasible. While D/D fusion is likely to be feasible in practice this question cannot be answered with certainty until easier D/T cavitation-induced fusion is studied in depth and appropriate bubble collapse and fusion models are constructed. This is one of the immediate tasks of CIF research. However, even if D/D fusion ends up being commercially unfeasible the D/D fusion can still be used for tritium production as tritium is one of the byproducts of D/D fusion. Therefore it may be possible to have commercial net-power producing reactors operating on D/T mixture while fuel-sourcing power-consuming reactors operating on D/D fuel will be used to produce tritium.

## 5. Preliminary Results

At the moment of writing our company has achieved the following preliminary results:

1. We have developed a numerical solution (Mathematica and MatLab) for supersonic bubble collapse governed by Rayleigh-Plesset-Keller equation (5) that accounts for acoustic losses due to shockwave compression and liquid compressibility. We can solve the equation (5) together with the deuterium equation of state (7) and estimate fusion reaction yield using the equations (9) – (13) to obtain a lower bound of fusion reaction rate since our calculations assume adiabatic collapse (uniform temperature and pressure) and ignore shockwave-related effects. Using this tool we were able to identify a range of parameters (such as maximum bubble radius, driving acoustic pressure, ambient bubble radius, liquid choice, etc.) that we will result in highest fusion probabilities in a laboratory setting.
2. We have modified molecular dynamics software originally developed at UCLA by Bass (Bass, Ruuth, Camara, Merriman, & Putterman, 2008) so we are able to simulate shockwave formation in the collapsing bubble – Fig. 13 – and calculate fusion reaction rate thus further increasing the accuracy of predictions obtained with the Mathematica and Matlab models and discovering a new range of parameters that has to do with shockwave formation, which is an extremely powerful mechanism for fusion initiation.
3. We have performed an initial series of cavitation-induced fusion experiments using the micro-reactor depicted on Fig. 14 and detected weak neutron emission (using Eberline ASP-1 $BF_3$ detector) coinciding with cavitation. The initial reactor design proved inadequate due to excessive power requirements and poor reaction yield stemming from inadequate resonator cavity design. Because fine-tuning the resonator is a laborious and costly process requiring precise numerical simulation that is sensitive to many design parameters (Lahey, Taleyarkhan, Nigmatulin, & Akhatov, 2006) we chose a new experiment design according to (Smorodov, Galiakhmetov, & Il'gamov, 2008), which is described in depth in the following section.
4. We have built a single-bubble fusion setup according to Smorodov (Smorodov, Galiakhmetov, & Il'gamov, 2008) – Fig. 15. In this experiment a single deuterium bubble is injected into a cylinder filled with chilled degased glycerin; the cylinder is capped with a tight-fitting piston, which creates a pressure wave when a weight is dropped on the piston. We are currently working on improving this experiment as the initial data was very encouraging: the first two experiments resulted in significant above-background neutron emission coincident with the impact (Eberline ASP-1 detector), however the subsequent six trials yielded no signal (in part because a leak developed within the system and we could not achieve the requisite pressures).
5. We have constructed a prototype of the commercial cavitation-induced fusion power generator, which relies on hydrodynamic cavitation for bubble generation – Fig. 16. The generator design is according to Kladov (Kladov, 1994) and the operating parameters will be configured according to the predictions of the molecular dynamics simulation.

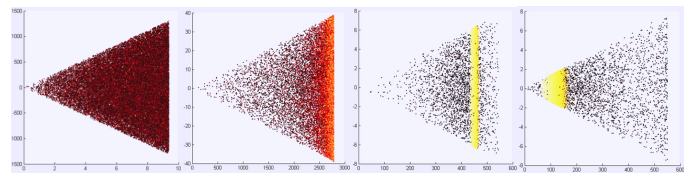

**Fig. 13.** Molecular dynamics simulation of shockwave formation within the collapsing cavitation bubble. In order to reduce the computational time we model only a conical section of the bubble.

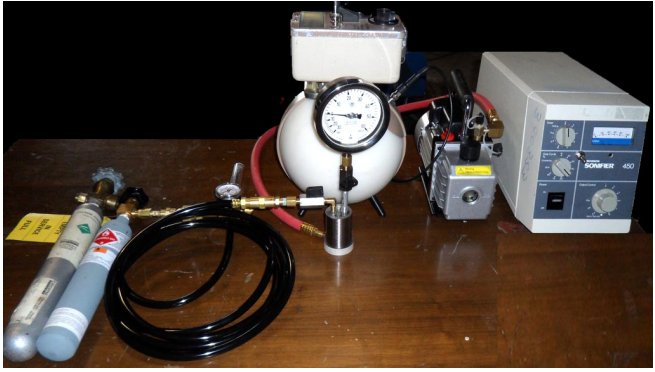

**Fig. 14.** Micro-reactor developed by our company for CIF experimentation: small cylindrical stainless steel reactor (middle) is fitted with a piezoelectric-ring (bottom of cylinder) driven by a power amplifier (right); pressure gauge is mounted on top of the chamber and connected to a vacuum pump; gas bottles (left) are connected to the chamber via a supply system; the reactor chamber is set in front Eberline ASP-1 $BF_3$ neutron detector enclosed in polyethylene moderator sphere.

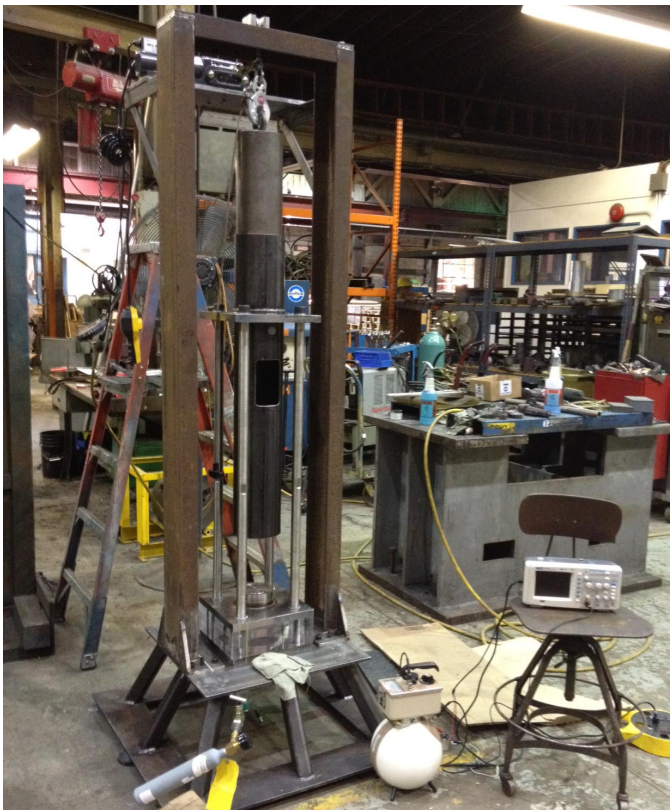

**Fig. 15.** The single-bubble cavitation-induced fusion experiment according to Smorodov (Smorodov, Galiakhmetov, & Il'gamov, 2008). After a deuterium bubble is injected into the glycerin the striker is released to impact the piston and create 300-500 bar pressure wave in the glycerin. The oscilloscope displays the signal from the neutron detector, which must coincide with the signal spike from the pressure transducer.

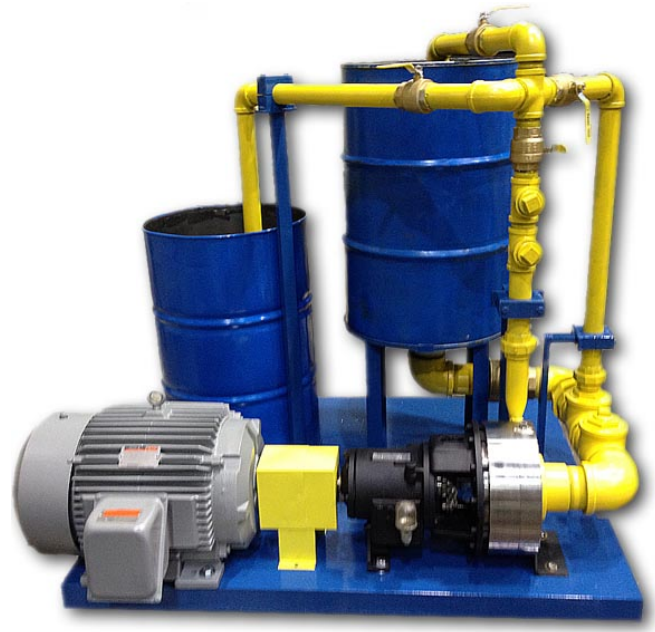

**Fig. 16.** Commercial 100-kW cavitation-induced fusion power generator prototype powered by 50HP electric motor.

## 6. Technical Proposal

### 6.1 Goal of Phase I

**The goal of the Phase I of the project is to show beyond any doubt that D/D or D/T fusion indeed occurs within the collapsing cavitation bubbles under the proposed conditions.** Publication of easily reproducible and convincing results will enable constructive peer discussion and research collaboration necessary for advancing CIF field of study as well as reestablish the field as a legitimate area of research.

### 6.2 Success Criteria

Current academic consensus is that that a convincing fusion experiment must satisfy the following criteria:

1. Neutron emission must be at a statistically significant above-background level;
2. Neutron spectrum must be consistent with the D/D or D/T fusion reaction (depending on the fuel mixture used);
3. Neutron counts must coincide with bubble collapse events, which can be detected either optically via a high-speed camera or acoustically by detecting a shockwave generated by rebounding bubbles;
4. There must be no other sources of neutron radiation in the laboratories that could confuse the results;

5. The experiment design must yield reproducible results that can be easily replicated by an independent third party group of investigators.

## 6.3 Technical Objectives

To achieve the goal set forth in Section 6.1 and to meet the success criteria described in Section 6.2 we must address the following technical objectives:

1. Develop hardware for bubble injection into a carrier liquid allowing for bubble radius control in the range of few micron to several mm;
2. Develop hardware for acoustically driving the injected bubbles via piezoelectric transducer (necessary for creating large bubbles with low gas pressure);
3. Develop hardware for imparting shock pressure into a carrier liquid in order to achieve peak pressures on the order of 1,000 bar;
4. Develop hardware for synchronizing the impact shock with the maximum expansion radius of the bubble (the bubble expanded acoustically via piezoelectric transducer must be crushed via compression shock when it reaches maximum radius);
5. Procure a high-efficiency neutron detector with spectrum resolution;
6. Develop data logging scheme where the neutron detector counts and spectrum information is synchronized with the bubble collapse event defined as pressure transducer voltage spike corresponding to the rebounding bubble shockwave.
7. Develop improved GPGPU and super-computer based modeling software to be used as an engineering tool in conjunction with our single-bubble experiment (Fig. 15) and for tuning of the prototype of the commercial CIF reactor (Fig. 16).

## 6.4 Experiment Design

To satisfy the technical objectives set forth in Section 6.3 we propose an experimental design shown on Fig. 17. This design was already partially realized in our single-bubble experiment (Fig. 15).

Past experiments by other researchers – such as (Taleyarkhan R. , West, Cho, Lahey, Nigmatulin, & Block, 2002) – are difficult to reproduce because they require expensive equipment and/or precisely manufactured resonator chambers (Lahey, Taleyarkhan, Nigmatulin, & Akhatov, 2006) thus failing to satisfy our reproducibility and ease of replication criterion (Success Criterion #5). Therefore our proposed experiment follows the design proposed by (Smorodov, Galiakhmetov, & Il'gamov, 2008).

The experimental hardware is comprised of metal cylinder (1) with sapphire ports (not shown) filled with a suitable carrier liquid such as glycerin (8). Glycerin is particularly well suited for the proposed experiment due to its high viscosity, extremely low vapor pressure, and high intensity of sonoluminescence, which is indicative of strong shockwaves lunched within the bubble gas.

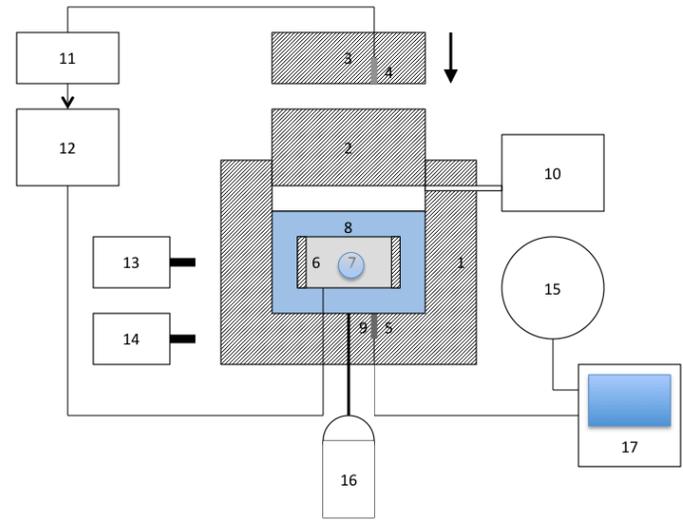

**Fig. 17.** The proposed experiment design to demonstrate feasibility of cavitation-induced fusion: 1 – Plexiglas cylinder, 2 – steel piston, 3 – weight to strike the piston, 4 – pressure transducer to trigger amplifier, 5 – hydrophone for shock monitoring, 6 – piezoelectric cylindrical transducer to drive bubble expansion, 7 – deuterium bubble suspended in glycerin, 8 – liquid glycerin, 9 – gas injection path, 10 – fore-vacuum pump, 11 – amplifier control circuit, 12 – amplifier to drive the piezoelectric transducer, 13 – high-speed camera, 14 – spectrometer, 15 – neutron counter, 16 – gas supply, 17 – digital storage oscilloscope.

The bottom section of the cylinder (1) contains a rubber-plugged channel (9) for injecting deuterium gas (or any other suitable gas mixture such as D/T or D/Xe, etc. that we wish to experiment with). The gas will be injected via a fine needle and we aim at being able to produce both micron and mm-size bubbles (7). Due to high viscosity of glycerin the bubbles will take a long time (tens of seconds) to rise to the top of the cylinder this giving us enough time to conduct our experiment.

Our calculations indicate that the highest fusion yield arises when the following conditions are met:

1. Bubble radius is as large as possible;
2. Driving pressure is as large as possible;
3. Bubble pressure is as small as possible;
4. Shockwave is lunched within the bubble.

Our options include injecting deuterium bubbles at atmospheric pressure or at reduced pressure, which can be achieved by removing air from the cylinder via a fore-vacuum pump (10). The lowest pressure we can achieve with our equipment is 50 Pa.

The other alternative is to inject a 10-micron deuterium bubble and oscillate it with a help of piezoelectric transducer ring (6) immersed in glycerin. By using modest acoustic driving pressure ($P_a \approx$ *few bar*) and low driving frequency ($f \approx$ *20 kHz*) we can achieve mm-size bubbles with low gas pressure without the need for chamber evacuation (the liquid may need to be degased prior to the experiment). To achieve fusion the maximum bubble radius needs to be synchronized with the impact of the piston (2). The synchronization can be accomplished via a pressure transducer (4) embedded in the weight (3) striking the piston. This signal from the transducer (4) will trigger the piezoamplifier (12) via the delay and control circuit (11) to direct the amplifier driving the piezoelectric transducer (6) to achieve the maximum bubble expansion just in time for peak pressure due to impact. The necessary delay in the amplifier operation will need to be established manually based on the experimentally determined impact pressure rise constant τ:

$$P(t) = P_{max} (1 - e^{-\tau/t}) \quad (15)$$

where

$$\tau = \frac{1}{2}\sqrt{\frac{m}{k}} \ln\left(\frac{2mgh}{k}\right) \quad (16)$$

To trigger the pressure transducer (4) just prior to the weight (3) impact we may need to pad the transducer head with a thin rubber sheet of experimentally determined thickness to get the transducer to trigger slightly prior to the weight impact.

At any rate we will first experiment with large bubbles at atmospheric pressure and move onto reduced pressures if our initial experiments are unsuccessful.

Once a bubble is formed in the liquid a heavy 50-kg weight (3) will be dropped on the piston (2) from maximum height $h_{max}$ = *1 m* to create a compression shock in the liquid. The peak impact pressure will be controlled by changing the drop height and will vary between $P_{peak}$ = *10 ÷ 2,000 bar*.

The reason we chose Plexiglas for cylinder material is because we want to monitor bubble behavior and sonoluminescence emission (if any) via a high-speed camera (13). This way we can track bubble collapse and detect shape instabilities (if any) as well as control bubble size during the injection stage.

To monitor fusion reactions we shall employ our ASP-3 $BF_3$ neutron detector/counter (15). The output of the detector and the hydrophone (5) will be fed to digital data logging oscilloscope (17) to determine correlation between the detector signal and the shockwave lunched by the rebounding bubble.

Additionally we shall employ four CR-39™ strips (Neutrak 144-T by Landauer) that are sensitive to fast, intermediate and thermal neutrons (energy spectrum of 0.5eV to 40 MeV; 0.20mSv minimum exposure). CR-39 plastic is a very robust neutron detector that cannot be influenced by electromagnetic radiation and electromagnetic noise, which is unavoidable in our experiment. Analytics services are readily available from Landauer with results given in mrem and neutron track counts (when neutrons strike plastic they create micro-tracks readily visible in microscope when inspecting CR-39 samples) will serve as an additional confirmation of neutron emission.

Lastly, we shall extract liquid samples after the experiment to conduct tritium detection assay to obtain yet another indicator of fusion reactions. Naturally, tritium assay will be performed only on deuterium filled bubbles and will not be used with D/T mixtures.

## 6. Related R&D

A comprehensive review of previous CIF-related research was given in [Section 3](). The research was conducted at the Russian Academy of Sciences, Oak Ridge National Laboratory, Los Alamos National Laboratory, Purdue University, and UCLA. At the moment of writing and as a fallout of "bubblegate" none of these institutions acknowledge active CIF research programs.

In the industry there is a sole privately funded company Impulse Devices, Inc. ([www.impulsedevices.com](http://www.impulsedevices.com)) that pursues research & development of Extreme Acoustic Cavitation™ with the objective to commercialize cavitation-induced fusion technology. The company was co-founded by Dr. Felipe Gaitan who is credited with the invention of single-bubble sonoluminescence, and now is a Chief Scientist at our company, Quantum Potential Corporation. Impulse Devices was a recipient of a $35-million dollar Advanced Cavitation Power Technology (ACPT) contract by DoD. Because it was a military contract no details or publications are available on the results of that work.

A Canadian startup company General Fusion ([www.generalfusion.com](http://www.generalfusion.com)) is pursuing a LINUS reactor design where a magnetically confined plasma is injected into a large cavity in molten lead that is driven to collapse by a synchronized action of 192 pistons sending a shock through the molten metal. General Fusion was funded at a $34M dollar level by the Canadian government and Canadian venture capitalists.

## 7. Conclusion

Because cavitation-induced fusion is viable and has been done before it is critical to restart active research in this area, which was abandoned for the wrong reasons. Smorodov's single-bubble fusion experiment and Kladov's hydrodynamic cavitation hardware are the best starting points in this quest. We have developed a three-year plan to develop a commercial CIF power generator and seek funding to complete the first year of work that will demonstrate convincingly repeatable, on-demand fusion that will reestablish the area as a legitimate field of research and pave the way to the commercial CIF power generation.